\documentstyle[12pt]{article}
\begin{document}

\title{New Functional Representation for Hubbard Model, Coherent States and Tower
of Algebras}
\author{{\large V. Zharkov} \\
{\em Institute of Natural Sciences, Perm State University,}\\
{\em Genkel st.4, 614600 Perm, Russia}}
\date{}
\maketitle

\begin{abstract}
New functional representation for the strongly interacting systems is
proposed which contains a new type of the coherent state. As a result the
new algebraic structure- so called "tower of algebras" appears which gives
the tower (or hierarchy) of models.
\end{abstract}

\section{Functional approach.}

Consider Hubbard model (lattice version of Thirring model) \cite{izum}:
\begin{equation}
H=-t\sum_{<{\bf r},{\bf r^{\prime }>,\sigma }}\,\alpha {}_{{\bf r\sigma }%
}{}^{+}\,\alpha {}_{{\bf r^{\prime }\sigma }}{}\,+\,U\sum_{{\bf r}}\,n{}_{%
{\bf r},\uparrow }n{}_{{\bf r},\downarrow }=H_{band}=
\end{equation}
\begin{equation}
\label{hubb}\sum_{{\bf r}}U\,X{}_{{\bf r}}{}^{22}+\sum_{A,C,{\bf r},{\bf %
r^{\prime }}}\,t{}_{-A}{}_C({\bf r}-{\bf r^{\prime }})\,X{}_{{\bf r}%
}{}^{-A}\,X{}_{{\bf r}^{\prime }}{}^C=H_{atomic}
\end{equation}
where $<r,r^{\prime }>$-denote the sum over the nearest neighbors. Electrons
are described by $(\alpha _{r\sigma }^{+},\alpha _{r\sigma })$ and live in
two dimensional plane $r=\{x,y\}$ (for example square lattice), $\sigma $ -
spin of the electrons. We give two different forms of this model (the first
form is a standard, the second contains the Hubbard operators $X^A$ (in fact
they are projectors $X_r^A=\mid ir><jr\mid $). These operators act in space
of following states:$\mid 0>,\mid +>,\mid ->,\mid 2>,\quad
A=(i,j);i,j=0,+,-,2$(the space of eigenfunctions of Hubbard repulsion-the
main and most complicated interaction in many interesting models). $X^{ij}$
has one nonzero element, sitting in i row and j column of $4\times 4$
matrix. Take $H_0=n_{\uparrow }n_{\downarrow }$ as zero field. Then
eigenfunctions and eigenvalues are:
\begin{equation}
\label{fun}
\begin{array}{cc}
\mid 0> & E_0=0 \\
\mid +>=a_{\uparrow }^{+}\mid 0> & E_{+}=-\mu  \\
\mid ->=a_{\downarrow }^{+}\mid 0> & E_{-}=-\mu  \\
\mid 2>=a_{\uparrow }^{+}a_{\downarrow }^{+}\mid 0> & E_2=U-2\mu
\end{array}
\end{equation}

here $\mu $-is chemical potential. Operators $a_\sigma ^{+},a_\sigma $ in
this base are:

\begin{equation}
\label{exp}a_\sigma ^{+}=X^{\sigma ,0}+\sigma X^{2,-\sigma }\quad a_\sigma
=X^{0,\sigma }+\sigma X^{-\sigma ,2}
\end{equation}

Local state of system and the Hilbert space of (\ref{hubb}) we describe by
supercoherent state:$\mid G[\chi (r,t^{\prime }),{\bf E}(r,t^{\prime }),{\bf %
h}(r,t^{\prime })>=Exp(-i{\bf E\rho }-i{\bf hs}-i{\bf \chi a})\mid 0>$ where
$\chi =\{\chi _1,\chi _2\}$ are dynamical two components odd valued
grassmann fields,${\bf E}=\{E_1,E_2,E_3\}$ and{\bf \ }$h=\{h_1,h_2,h_3\}$
are three components even valued dynamical fields. $\chi $ describes
electronic degrees of freedoms, ${\bf E}$ and ${\bf h}$ parametrize charge
density{\bf \ }${\bf \rho }$ and spin density{\bf \ }${\bf s}$ fluctuations
consequently. According to this formulation the many particles system are
described by following effective functional \cite{zhar84}:
\begin{equation}
\label{act}L_{eff}={\frac{{<\,G({\bf r},t^{\prime })\,\mid }\,\left( {\frac{{%
\partial }}{{\partial t^{\prime }}}}-H\right) \,\mid G({\bf r},t^{\prime
})\,>}{{<\,G({\bf r},t^{\prime })\,\mid \,G({\bf r},t^{\prime })\,>}}}
\end{equation}
where $\mid \,G\,>$ -is the supercoherent state, which is expressed through
generators of the dynamical superalgebra, $\{{\bf r},t^{\prime }\}$
-coordinates of space-time. H is the Hubbard hamiltonian expressed through
the infinite dimensional superalgebra generators $\{\,X_{{\bf r}}^\alpha \,\}
$(atomic representation): where $U-$ on-site coulomb repulsion, $t_{\alpha
,\beta }({\bf r-r^{\prime }})$ - ''interaction'' arisen from kinetic energy.
$\mid \,G\,>$ can be constructed by the following expression:

$$
\mid G>=Exp\left[
\begin{array}{cccc}
E_z & 0 & 0 & E^{+} \\
\chi _1 & h_z & h^{+} & 0 \\
\chi _2 & h^{-} & -h_z & 0 \\
E^{-} & -\chi _1 & \chi _2 & -E_z
\end{array}
\right] \mid 0> =( F \chi, Z(E)+2B \chi_{2} \chi_{1} )
$$

$(h_1,h_2,h_3,E_1,E_2,E_3,\chi _1,\chi _2)$ is the set of bose and fermi
fields.

\begin{equation}
\label{1}F=a^{\prime \prime }+a^{\prime }\,E_z+{\bf h}\,{\bf \sigma ^P}%
\,\left( a^{\prime }+a\,E_z\right) ,\quad B=\left( \{0,a\}+\rho (\psi
)\,z(\tau )\,E{}^{+}\right)
\end{equation}

where the prime means differentiation with respect to $\delta $,\thinspace $%
\sigma _\mu ^P$ -Pauli matrix.
$$
z(\tau )=\{\cos (\tau )+\cos (\theta ^{\prime })\,\sin (\tau ),{e^{i\,\phi
^{\prime }}}\,\sin (\tau )\,\sin (\theta ^{\prime })\},\tau ={\it \arcsin }({%
\ \frac{{E\,\psi }}\rho }),
$$
$$
\rho ={\sqrt{{E^2}\,{{\psi }^2}+{{\psi ^{\prime }}^2}},}\qquad \psi ={{%
\delta }^5}\,{\frac{{\partial \,f\,}}{{\partial \,E^2}}},a={\frac{{{{%
\partial }^2f}}}{{{E^2}\,{{\partial \delta }^2}}}}
$$
\begin{equation}
\label{2}\qquad f=-{h^{-2}}+{\frac{{{E^2}\,\left( {\frac{{\sin (E\delta )}}{{%
E^3}}}-{\frac{{\sin (h\delta )}}{{h^3}}}\right) }}{{{E^2}-{h^2}}}}
\end{equation}
We use the spherical coordinate system $\{\,E,\theta ^{\prime },\phi
^{\prime }\,\}$ for {\bf E}. We put $\delta =1$ in (\ref{1},\ref{2}) after
calculation. The partition function of the system now can be written as:
$$
Z_{hub}=\int D\chi {}^{*}\,D\chi \,D{\bf E}\,D{\bf h}\,{e^{-i\int
d{}^2r\,dt^{\prime }\,L({\bf E},{\bf h},\chi )}}
$$
Introduce so called QP-derivative: $D_{QP}K(x)=(K(Px)-K(Qx))/((P-Q)x)$.
Using this derivative one can represent:

$K(x)=ch(\sqrt{\alpha })/\alpha ;~D_{E^2,h^2}K(\alpha )=(ch(\sqrt{\alpha }%
E)/(\alpha E^2)-ch(\sqrt{\alpha }h)/(\alpha h^2))/(\alpha (E^2-h^2))=f(\sqrt{%
\alpha })/\alpha ^2$

It follows here from that all expressions are given as action of
two-parameters deformed derivative. As a result we have the appearance of
quantum algebra. Our main task will be the investigation of these deformed
algebraic structure. Consider several possibility: a) $E_k=0,h_k=0$, then

$$
\mid G>=\left(
\begin{array}{c}
1 \\
\chi _1 \\
\chi _2 \\
2\chi _2\chi _1
\end{array}
\right)
$$

this expression coincide with fermionic CS; b)$\chi =0,h =0$ then $\mid
G>=Z(E)$ and density operators give us $\rho _3\Rightarrow
z_0^{*}z_0-z_2^{*}z_2;~\rho ^{-}\Rightarrow z_2^{*}z_0,~\rho ^{+}\Rightarrow
z_0^{*}z_2.$ The set of $\rho _1,\rho _2,\rho _3$ gives some algebra. This
is a CS for SU(2) if $E=\pi /2,3\pi /2,...$. In this case $\rho
_3\Rightarrow cos(2\theta )$, $\rho ^{-}\Rightarrow e^{i\phi }sin(2\theta
),~\rho ^{+}\Rightarrow e^{-i\phi }sin(2\theta ).$ $E$ parameterises Casimir
operator and gives the value of quasispin. In interval $0<E<\pi /2$ this is
the quantum coherent state for $SU_q(2)$. For spin operator we have $SU_q(2)$
$$
S_q^{\pm }=h^{\pm }\frac{\sin h}h(cos(h)\pm h_z\frac{\sin h}h);\quad
S_q^z=cos^2h+\frac{\sin ^2h}{h^2}(h_z^2-h^{+}h^{-})
$$

Evaluation of Poisson brackets gives us deformed SU(2) algebra.Thus CS gives
two parameters deformed $SU(2)*SU(2)$ (we ignore the difference between
SU(2) and SU(1,1) for density subsystem), where $h,E$ (magnetic moment value
and density parameter (bandwidth)). Let's evaluate the expressions for
symbols:

$$
\begin{array}{c}
\rho _3\Rightarrow Z^{*}s_zZ+2(Z^{*}s_zB)\chi _2\chi _1+2(B^{*}s_zZ)\chi
_1^{*}\chi _2^{*}+4(B^{*}s_zB)\chi ^{*2}\chi ^2 \\
\rho ^{-}\Rightarrow z_2^{*}z_0+2z_0B_2^{*}\chi _2\chi _1+2z_2^{*}B_0\chi
_1^{*}\chi _2^{*}+4B_0^{*}B_2\chi ^{*2}\chi ^2 \\
\rho ^{+}\Rightarrow z_0^{*}z_2+2z_0^{*}B_2\chi _2\chi _1+2z_2B_0^{*}\chi
_1^{*}\chi _2^{*}+4B_2^{*}B_0\chi ^{*2}\chi ^2
\end{array}
$$
Introduce for the spin symbol :$S^{\prime }=S/(a^{\prime 2}-a^2E_3^2)$
$$
\begin{array}{c}
S_z^{\prime }\rightarrow Q_z^hs_z-h_zIm(m_0)\rho
_3+h^{+}(Re(m_0)-2h_z)s^{-}-h^{-}(Re(m_0)+2h_z)s^{+} \\
S^{\prime +}\mapsto [-Im(m_0)h^{+}\rho
_3-h^{+}h_zs_z-h^{+2}s^{-}+(m_o^{*}+h_z)(m_o+h_z)s^{+}] \\
S^{\prime -}\rightarrow [-Im(m_0)h^{-}\rho
_3-h^{-}h_zs_z-h^{-2}s^{+}+(m_o^{*}-h_z)(m_o-h_z)s^{-}]
\end{array}
$$
here $Q_z^h=m_0^{*}m_0-h_z^2+h^{+}h^{-}.$ Symbols for fermionic operators: $%
<g\mid a_{+}\mid g>/<g\mid g>=z_0^{*}(F\chi )_1+z_2(F\chi )_2^{*}$. Consider
the following cases: $E\rightarrow 0,h\rightarrow 0$, thus

$F\rightarrow \left(
\begin{array}{cc}
F_{11} & 0 \\
0 & F_{22}
\end{array}
\right) ;$

and for symbol of creation-annihilation operators one can get:

$a_\sigma \rightarrow F_{\sigma \sigma }\chi _\sigma ;\quad B\rightarrow
(0,b);$

As a result we arrive to the canonical grassmann representation for
fermionic systems: $a_\sigma \rightarrow \chi _\sigma $. Let's obtain the
representation for spin. Taking mean value for the density operator one can
obtain the expansion for spin:

$S^{+}=(1-\alpha )S_q^{+}+\alpha \chi _1^{*}\chi _2;\quad S^{-}=(1-\alpha
)S_q^{-}+\alpha \chi _2^{*}\chi _1;\quad S^z=(1-\alpha )S_q^z+\alpha (\chi
_1^{*}\chi _1-\chi _2^{*}\chi _2);$

The set $(\chi _1^{*}\chi _2,\chi _2^{*}\chi _1,\chi _1^{*}\chi _1-\chi
_2^{*}\chi _2)$ is a classical SU(2) algebra. This expansion indicates that
localized moments are defined as the sum of two parts: classical and quantum
so that the sum of these two parts equal to unity, parameter $1- \alpha $ is
the order parameter of magnetic localization.

{\bf TOWER OF ALGEBRAS}: Consider the expression for $a_\sigma \sim a\chi
+b(E+h)\chi +cEh\chi =A_0+A_1+A_2$ . It is evident that in enveloping
algebra there possible only polynomials of three types $A_0,A_1,A_2:$1)
fermionic sector of $A_1$ contains 4 odd operators, 2) fermionic sector of $%
A_2$ contains 8 odd operators; 3) fermionic sector of $A_0$ contains 2 odd
operators.Terms in expansion describe: metal - insulator - spinless
noninteracting gas algebras of variables. Thus we have some expansion for $%
a_\sigma $ through bases of 3 different algebras. This expansion was called
''tower of algebras'' \cite{zhar92}.

{\bf \ METAL-INSULATOR TRANSITION: }Consider those deformations which do not
change a metallic algebra: $a_{\uparrow }=z_1^{*}(F\chi )_1-z_2^{*}(F\chi
)_2;\quad a_{\downarrow }=z_1^{*}(F\chi )_2+z_2^{*}(F\chi )_1;$ Saddle point
approximation for energy gives us: $%
E_H=-(F_{11}^{*}F_{11}+F_{22}^{*}F_{22})+b^2U;$ Evaluation of this
expression gives: $E_H=-(E^2+h^2)[sin^2E+sin^2h-b^2]+b^2U$ . Expanding
energy on the parameters $E<<1,h<<1$ we arrive to Gutzwiller approximation :
$E_H=-\nu (1/2-\nu )+24\nu U$ \qquad here $\nu =E^2+h^2<<1$ . We obtain the
following interpretation for $\nu $ -it is a combination of two deformation
parameters.

\section{ Operator approach}

Let's give the operator's version of tower of algebras and tower of models.
We start from Hubbard model in the atomic representation :
\begin{equation}
\label{huba}H= \sum_{r,p}E_pX^{pp}+ \sum_{\alpha ^{\prime}\beta ,rr^{\prime
}} t^{-\alpha ^{\prime }\beta }(r-r^{\prime })X_r^{-\alpha ^{\prime
}}X_{r^{\prime }}^\beta
\end{equation}
In atomic base $a_\sigma $ have following nonzero matrix elements:$<0\mid
a_{\uparrow }\mid +>=<0\mid a_{\uparrow }\mid +>=1$. One can write the
following identity:
\begin{equation}
\label{exp1}Op_{\uparrow }=(1-\alpha )a_{\uparrow }+\alpha (X^{0\uparrow
}-X^{\downarrow 2})
\end{equation}
here $Op_{\uparrow }$ -operators for which: $\{Op_\sigma ^{+},Op_{\sigma
\prime }\}=\delta _{\sigma ,\sigma \prime }$ and $0<\alpha <1$.

Let's take $a^{+},a$ for which $\{a^{+},a\}=1$, for example: ($a\sim
a_\sigma +a_{-\sigma }$). We can construct superalgebra: $%
A_0=(a^{+},a,n=a^{+}a)$. Take $a,a^{+}$ and construct the following
operators: $(\frac a2(\gamma _2+s_z)+(\gamma _2-s_z)\frac a2,\frac a2(\gamma
_2-s_z)+(\gamma _2+s_z)\frac a2)$, here $\gamma _3=1-n;\gamma _2=1-(\gamma
_3)^2;s_z=diag(0,1,-1,0).$ It is seen that: 1) this set is algebra $A_1$; 2)
the operators are polynomials on $a,a^{+}$. Number of fermionic generators
equal 4. The third step: taking $\gamma ^5=diag(1,-1,-1,1)$ and $(a_\sigma
^{+},a_\sigma ,{\bf {\rho }},{\bf {s}})$ one can construct polynomials
of $a,a^{+}$ (take chiral projection by multiplying $1\pm \gamma ^5$), which
gives algebra $A_2$ ( $X^{pq}\sim \gamma ^5a_\sigma $; $\gamma ^5\sim \rho
^2-s^2$, thus $X\sim a^5$). The set of $A_0,A_1,A_2$ as linear space gives
the universal enveloping of $A_0$. Let's construct from $A_0,A_1,A_2$ a new
set of variables:
\begin{equation}
\label{ealg}a_\sigma ^{\prime }=(1-\alpha )a_\sigma ,X^{\prime pq}=\alpha
X^{pq},
\end{equation}
where $X^{pq}$ -fermi operator,
$$
\{X^{\prime +-},X^{\prime -+},X^{\prime z},X^{\prime 02},X^{\prime
20},X^{\prime 00}-X^{\prime 22}\}
$$
here $X^{\prime pq}=X^{pq}$.

(Anti)commutation relations for ($a^{\prime },X^{\prime }$) are:
\begin{equation}
\label{com}\{a_\sigma ^{\prime },a_{\sigma ^{\prime }}^{\prime
,+}\}=(1-\alpha )^2\delta _{\sigma \sigma ^{\prime }}\quad [X^{\prime
pq},X^{\prime qQ}\}=\alpha ^2X^{\prime pQ}\quad [X^{\prime +-},X^{\prime
-+}]=X^{\prime ++}-X^{\prime --}
\end{equation}
$$
\{a_{\uparrow }^{\prime },X^{\prime +0}\}=\alpha (1-\alpha )X^{\prime ++}
$$
Structural constants are the functions of $\alpha $. When $\alpha =0$ we have
$A_1$. When $\alpha =1$ , we have insulator algebra $A_2$. Varying $\alpha $
we can obtain the interpolation between metallic and insulator algebras. As
a result we can describe two level (or places) of tower. Using these two
level algebra one can obtain expression for hamiltonian:
\begin{equation}
\label{hubort}H=(1-\alpha )^2H_{band}+\alpha ^2H_{atomic}+\alpha (1-\alpha
)H_{band}^{atomic}+\alpha (1-\alpha )H_{atomic}^{band}
\end{equation}
It is two-band model: the band $a_\sigma $ describes delocalized electrons,
the second band $X^{pq}$ describes atomic states.Thus we obtain from the
one-band model an effective two-band model. If $\alpha =0$ it is
weakly-interacting gas, when $\alpha =1$ we have localized insulator. We
have the possibility to describe the following hierarchy of model: spinless
gas- fermi liquid - magnetic insulator - two-band model. From the point of
view of functional integral we have the following chain (or tower) of
dynamical algebras defining local bosonic variables: $U(1)\rightarrow
U(2)\rightarrow U(4)\rightarrow U(8)$ . It is obvious that the last
structure which is based on $U(8)$ local algebra is very rich from field
theoretical point of view. Let's give physical view of functional integral
for strongly interacting systems in which the tower of algebras appears. The
stabilization of tower means that on the lattice scale we have quantum
fluctuations between many ground states described by the tower of models (or
the infinite dimensional algebra (quantum many phase state) which contains
the infinite number of subalgebras of concrete model of definite ground
state - metal, insulator, magnet, superconductor ). It is a high energy
regime (the analog of UV regime in the field theory). At medium range of
energy as a result of the spontaneous breaking of scale invariance definite
dynamical algebra is stabilized and as a result a concrete type of the
ground state become stable. In this region the local variables are described
by the quantum groups. At the limit of low energy (IR regime in the field
theory) we have scenario of the spontaneous breaking of usual (the
classical) symmetry. As a result we have very rich structure of the
functional integral (the structure of measure, renormalization group,
effective action, the methods of evaluation ). This work in part is
supported by RFFI under Grant N. 95-03-08287.

\end{document}